\documentclass[12pt]{article}

\usepackage{latexsym} 
\usepackage{amssymb}  
\usepackage{amsbsy}   
\usepackage{graphicx}       
\usepackage[dvips]{color}
\usepackage{bm}   

\newcommand{\al}{\alpha}
\newcommand{\be}{\beta}
\newcommand{\ga}{\gamma}
\newcommand{\Ga}{\Gamma}
\newcommand{\de}{\delta}
\newcommand{\De}{\Delta}

\newcommand{\eps}{\epsilon}

\newcommand{\la}{\lambda}

\newcommand{\si}{\sigma}

\newcommand{\om}{\omega}

\newcommand{\<}{\langle} 
\renewcommand{\>}{\rangle} 
\newcommand{\txt}{\textstyle}
\newcommand{\dsp}{\displaystyle}
\newcommand\eqn[1]{(\ref{#1})}      
\newcommand\Eqn[1]{Eq.~(\ref{#1})}  
\newcommand{\e}{{\rm e}}   
\newcommand{\beq}{\begin{equation}}
\newcommand{\eeq}{\end{equation}}
\newcommand{\ba}{\begin{array}}
\newcommand{\ea}{\end{array}}
\newcommand{\bea}{\begin{eqnarray}}
\newcommand{\eea}{\end{eqnarray}}
\newcommand{\bi}{\begin{itemize}}  
\newcommand{\ei}{\end{itemize}}
\newcommand{\ben}{\begin{enumerate}} 
\newcommand{\een}{\end{enumerate}}
\newcommand{\bc}{\begin{center}}
\newcommand{\ec}{\end{center}}

\newcommand{\half} {{\txt \frac{1}{2}}}
\newcommand{\third}{{\txt \frac{1}{3}}}

\newcommand{\twothirds}{{\txt \frac{2}{3}}}

\makeatletter 

\def\appendix{\par                              
    \setcounter{section}{0}                     
    \setcounter{subsection}{0}
    \renewcommand{\theequation}{\Alph{section}.\arabic{equation}}
    \renewcommand{\thesection}{Appendix \Alph{section}
                \setcounter{equation}{0}  } 
    \renewcommand{\thesubsection}{\Alph{section}.\arabic{subsection}}
}

\def\applabel#1{\@bsphack
  \protected@write\@auxout{}%
         {\string\newlabel{#1}{{\Alph{section}}{\thepage}}}%
  \@esphack}

\def\section{
\setcounter{equation}{0}        
\@startsection {section}{1}{\z@}{-3.5ex plus -1ex minus 
 -.2ex}{2.3ex plus .2ex}{\large\bf}}
\renewcommand{\theequation}{\arabic{section}.\arabic{equation}}

\def\subsection{\@startsection{subsection}{2}{\z@}{-3.25ex plus -1ex minus 
 -.2ex}{1.5ex plus .2ex}{\normalsize\bf}}

\def\subsubsection{\@startsection{subsubsection}{3}{\z@}{-3.25ex plus
 -1ex minus -.2ex}{1.5ex plus .2ex}{\normalsize}}

\makeatother   

\newcommand{\eV}{{\rm eV}}
 
\newcommand{\MeV}{{\rm MeV}}

\newcommand{\Qt}{{\tilde Q} }

\newcommand{\diag}{{\rm diag}}

\newcommand{\one}{\bm{1}}
\newcommand{\dmu}{\de\mu}

\newcommand{\mubar}{{\bar\mu}}
\newcommand{\openone}{\bm{1}}

\usepackage[normalem]{ulem}  
\newcommand{\nova}[1]{#1}

\begin{document}

 
\title{\bf Photons
in gapless color-flavor-locked quark matter}


\author{
Mark Alford and Qinghai Wang \\[2ex]
Department of Physics \\ Washington University \\
St.~Louis, MO~63130 \\ USA}

\date{April 6, 2005}

\begin{titlepage}
\maketitle
\renewcommand{\thepage}{}          

\begin{abstract}
We calculate the Debye and Meissner masses of a gauge boson in a material
consisting of two species of massless fermions that form
a condensate of Cooper pairs. We perform the calculation
as a function of temperature, for the cases of neutral Cooper pairs
and charged Cooper pairs, and for a range of parameters including
gapped quasiparticles, and ungapped quasiparticles with both
quadratic and linear dispersion relations at low energy.

Our results are relevant to the behavior of photons and gluons in
the gapless color-flavor-locked phase of quark matter. We find that the
photon's Meissner mass vanishes, and the Debye mass
shows a non-monotonic temperature dependence, and 
at temperatures of order the pairing gap it drops to
a minimum value of order $\sqrt{\alpha}$ times the
quark chemical potential. We confirm previous claims that
at zero temperature an imaginary Meissner mass can arise
from a charged gapless condensate, and we 
find that at finite temperature this can also occur
for a gapped condensate.
\end{abstract}

\end{titlepage}


\section{Introduction}
\label{sec:intro}

In this paper we calculate the Debye and Meissner masses of a gauge
boson propagating through a material
consisting of two species of massless charged spin-$\half$ fermions.
We assume that,
via an unspecified pointlike attractive interaction, these form an
$s$-wave (rotationally invariant) condensate of Cooper
pairs, which may be neutral or charged depending on the charges
of the fermions. We allow the two species to have chemical potentials
$\mubar\pm\dmu$, and pairing gap parameter $\De$, which is
momentum-independent because the interaction is pointlike.
As one varies the
chemical potential splitting $\dmu$ of the two species, the
spectrum of fermionic excitations (quasiparticles) changes
dramatically: 
\beq
\ba{l@{\qquad}l@{\qquad}l}
\De > \dmu & \mbox{Gapped spectrum} 
  & \mbox{e.g.: Fig.~\ref{fig:disprel}, dotted line} \\
\De = \dmu & \mbox{Ungapped quadratic spectrum} 
  & \mbox{e.g.: Fig.~\ref{fig:disprel}, solid line} \\
\De < \dmu & \mbox{Ungapped linear spectrum} 
  & \mbox{e.g.: Fig.~\ref{fig:disprel}, dashed line} \\
\ea
\label{disprel_cases}
\eeq
We calculate the zero-momentum current-current correlation function
(i.e.~the Meissner and Debye masses of the corresponding gauge boson)
for all these cases, for both a charged and a neutral condensate, as
a function of temperature.

To explain the motivation for this calculation, 
we give in section \ref{sec:gCFL} 
a summary of the properties of ultra-dense quark matter, focussing in
particular on the gapless color-flavor-locked (gCFL)
phase \cite{gCFL}, in which all these different cases
occur\footnote{The strong interaction that drives the quark
pairing is not pointlike, so the gCFL gap parameters are not
momentum-independent. However, we do not expect our results to be sensitive
to this feature.}. Our calculation is described in sections
\ref{sec:quasiparticles} and \ref{sec:self-energy}, with technical details 
in an appendix. The results are presented and discussed in sections
\ref{sec:neutral} and \ref{sec:charged}, with concluding remarks in
section \ref{sec:conclusions}.

\section{Gapless color-flavor-locked quark matter}
\label{sec:gCFL}

\subsection{Overview of the gCFL phase}

The behavior of matter at ultra-high density 
has been the subject of much theoretical work, and is gradually being
constrained by experiments. It is generally agreed that at sufficiently
high density, matter will be in a color-flavor-locked (CFL)
color-superconducting quark matter phase \cite{CFL} (for reviews,
see Ref.~\cite{Reviews}). However, there are
major questions about the next phase down in density. Recent
work \cite{gCFL} suggests that when the density drops low enough so that
the mass of the strange quark can no longer be neglected, there
is a continuous phase transition from the CFL phase to a new
gapless CFL (gCFL) phase. The two are very different: CFL quark matter
is a transparent insulator, with no electrons \cite{CFLneutral}. 
Its only light mode is a neutral
superfluid mode associated with the breaking of baryon number.
In contrast, gCFL quark matter is more like a metal: it has
gapless quark modes, and electrons, as well as the superfluid mode.

There are many unanswered theoretical questions about gCFL quark matter.
It may be modified by condensation of ``kaons'' \cite{gCFL-K0}.
Also, initial calculations indicate that some of the gluons have imaginary
Meissner masses, indicating an instability towards the development
of color currents \cite{Huang:2004am,Casalbuoni:2004tb,Giannakis:2004pf}. 
Our results (section \ref{sec:charged}) confirm that imaginary Meissner 
masses are associated with gapless charged condensates.
\nova{If the gCFL phase turns out to be stable after all, then
an experimental question will arise: how could one}
detect gCFL quark matter in nature. The
best opportunity for this is in the core of a neutron star, which achieves
densities well above nuclear density, at low temperatures that
permit color superconductivity. As we will discuss below, the most 
characteristic properties of gCFL quark matter are its transport
properties. For example, the presence of
a gCFL region will have a strong effect on the cooling of a neutron star
\cite{Alford:2004zr}. Further progress will require the calculation
of the interactions among the lightest excitations, and in this
paper we will lay the groundwork for such investigations by performing
a very general calculation of the behavior of zero-momentum gauge bosons
coupled to both neutral and charged Cooper pair condensates.

\begin{figure}[tb]
\begin{center}
\includegraphics[width=0.6\textwidth]{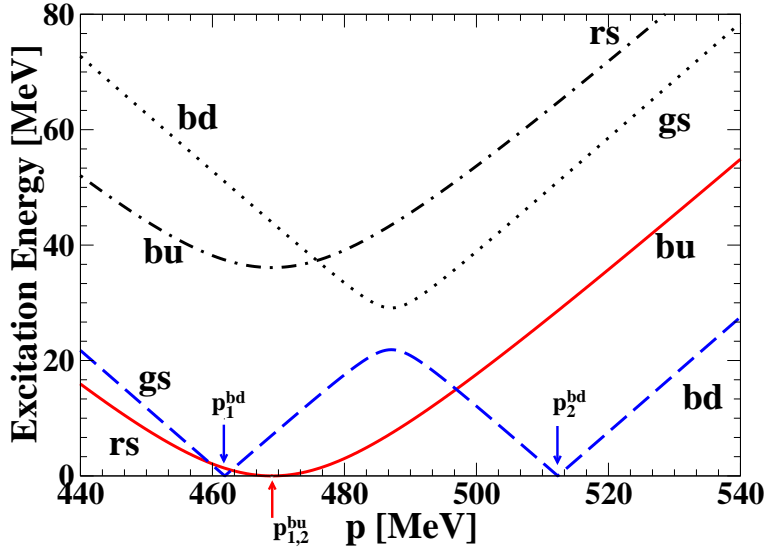}
\end{center}
\caption{\small
Dispersion relations for some of the quasiquarks in the gapless CFL phase.
Note that there is a gapless mode with an approximately
{\em quadratic}\/ dispersion relation (solid line)
as well as two gapless modes with more conventional linear dispersion
relations (dashed line), and gapped modes. The labels give the color ($r,g,b$)
and flavor ($u,d,s$) content of the quasiparticles.
}
\label{fig:disprel}
\end{figure}

\subsection{Quasiparticle dispersion relations in the gCFL phase}

In the gCFL phase, there is a condensate of Cooper pairs of quarks
in the color-antisymmetric, flavor-antisymmetric, and Dirac-antisymmetric
channel (there is also a insignificant color-symmetric flavor-symmetric
component, which we neglect)
\beq
\langle \psi^\alpha_a C\gamma_5 \psi^\beta_b \rangle \sim 
\Delta_1 \eps^{\alpha\beta 1}\eps_{ab1} \!+\! 
\Delta_2 \eps^{\alpha\beta 2}\eps_{ab2} \!+\! 
\Delta_3 \eps^{\alpha\beta 3}\eps_{ab3} \ .
\label{gCFL_condensate}
\eeq
Here $\psi^\alpha_a$ is a quark of color $\alpha=(r,g,b)$ 
and flavor $a=(u,d,s)$.  The gap parameters
$\De_1$, $\De_2$ and $\De_3$ describe down-strange,
up-strange and up-down Cooper pairs, respectively. 

The gCFL phase is named after its most striking characteristic:
the presence of gapless modes in the spectrum of quark excitations
above the color-superconducting ground state \eqn{condensate}. 
These ``gapless quasiquarks''
were discussed in Ref.~\cite{gCFL}, and a sample dispersion relation plot
is shown in Fig.~\ref{fig:disprel}. This was calculated using a
Nambu-Jona-Lasinio model of the strong quark-quark interaction,
with a quark chemical potential $\mubar=500~\MeV$,
strange quark mass  $m_s=200~\MeV$ and pairing strength chosen such that
at $m_s=0$ the CFL gap would be $\De_0=25~\MeV$ \cite{gCFL}.
There are also gapped quasiquarks whose dispersion relations are not shown.

The surprising feature of the gCFL phase is that there is a gapless mode
with an approximately quadratic dispersion relation $E(p) \propto (p-p_F)^2$,
as well as gapless modes with the more typical linear dispersion relation
$E(p) \propto |p-p_F|$, and fully gapped modes. The quasiquark spectrum
therefore includes all the cases listed in \eqn{disprel_cases}.
The presence of the quadratic gapless mode
means that there is a lot of phase space at low energy, in fact 
the density of states for the quadratic mode diverges as $E\to 0$.
We expect this to have dramatic effects on the transport properties,
and it has already been pointed out that gCFL quark matter has
an unusually large specific heat, which
varies as $c_V\propto\sqrt{T}$ rather than the usual $c_V\propto T$
that is associated with linear gapless dispersion relations.
This will alter the late-time cooling of a neutron star in an
observable way \cite{Alford:2004zr}. To make further progress in
studying the transport properties, it will be necessary to understand
the dominant (electromagnetic) interactions of the lightest quasiparticles.
For this we need the in-medium properties of the photon, which is why 
calculations of the Debye and Meissner mass of the photon are important.

\subsection{Relating our two-species calculation to the gCFL phase}

Our calculation explores the case of two species of fermions. This
might seem to be a drastic simplification of the gCFL phase, in which
there are nine species, but actually the gCFL pairing pattern
decomposes into three sectors of two-species pairing and one
of three-species pairing. Once pairing has occurred, the
dominant interaction between the quasiparticles is mediated by the
gauge boson of the remaining unbroken $U(1)_\Qt$ gauge symmetry, whose
gauge boson is mostly the original photon, with a small admixture
of one of the gluons. The relevant quality of the quasiquark excitations
is therefore their $\Qt$ charge, given by
$\Qt = \diag(\twothirds,-\third,-\third)_{\rm flavor}
- \diag(\twothirds,-\third,-\third)_{\rm color}$,
not their electromagnetic charge.
The pairing pattern and $\Qt$-charges of the quasiparticles are given in
Table~\ref{tab:charges}. Each two-species sector has an average
chemical potential $\mubar$ and a splitting $\dmu$ that arises from
the constraint of electrical neutrality and an approximate treatment
of the strange quark mass, in which it is treated as a contribution
$-M_s^2/(2\mubar)$ to the chemical potential for strangeness.
This is known to be a good 
approximation \cite{Fukushima:2004zq,Shovkovy:2004um}.

\begin{table}
\newcommand{\mystrut}{\rule[-1.5ex]{0em}{4ex}}
\newlength{\cellwid}
\settowidth{\cellwid}{gapless}
\begin{tabular}{lccccccccc}
\hline
species   \mystrut
  & ~$ru$~ & ~$gd$~ & ~$bs$~~~
  & ~~~$gu$~ & ~$rd$~~~
  & ~~~$rs$~ & ~$bu$~~~
  & ~~~$gs$~ & ~$bd$~~~
  \\
$\Qt$-charge   \mystrut
  & 0 & 0 & 0
  & $+1$ & $-1$
  & $-1$ & $+1$
  & 0 & 0
  \\
gap parameter   \mystrut
  & \multicolumn{3}{c}{$\De_1,\De_2,\De_3$}
  & \multicolumn{2}{c}{$\De_3$} 
  & \multicolumn{2}{c}{$\De_2$}
  & \multicolumn{2}{c}{$\De_1$}
  \\
quasiparticles   \mystrut
  & \multicolumn{3}{c}{gapped}
  & \multicolumn{2}{c}{gapped}
  & \multicolumn{2}{c}{\parbox{\cellwid}{gapless\\ (quadratic)}}
  & \multicolumn{2}{c}{\parbox{\cellwid}{gapless\\ (linear)}}
  \\
\hline
\end{tabular}
\caption{The structure of the gCFL pairing pattern, which decomposes into
one sector of three mutually paired species, and three sectors of two
species that pair with each other. The $\Qt$ charges of the quarks determine 
their electromagnetic interactions. Our two-species calculation can be applied
to the $gu$-$rd$, $rs$-$bu$, and $gs$-$bd$ sectors.
}
\label{tab:charges}
\end{table}

From Table~\ref{tab:charges} we see that the transport properties
will be dominated by the only electromagnetically interacting
light degrees of freedom, namely the $bu$/$rs$ quasiquarks
(all the others are either gapped or $\Qt$-neutral) and the electrons.
At any nonzero temperature
there will be a nonzero density of these $\Qt$-charged particles
which may lead to screening of the $\Qt$-electromagnetic
fields. Because of their divergent density of states at low energy,
we expect the $bu$-$rs$ quasiparticles to dominate this screening.
The most direct application of the calculations in this paper
is therefore to determine the in-medium properties of the photon
in gCFL matter, taking into account the effect of the $bu$/$rs$ quasiquarks.
Since the $bu$ and $rs$ quarks have opposite charge, the relevant results
are those for a neutral condensate, presented in section \ref{sec:neutral}.

We should note, however, that our calculations of Debye and Meissner
masses for charged condensates are also relevant: they can shed light
on the behavior of the gauge bosons associated with the broken 
$SU(3)_{\rm color}$ generators,
the gluons\footnote{The eighth gluon mixes with the photon, so
the corresponding broken generator is also associated with 
a photon-gluon mixture. The other broken generators are entirely gluonic.}.
The diagonal color generators can be treated as $U(1)$ gauge fields
with their own charge assignments to the quarks, different from the
$\Qt$ charges. The fact that they are broken corresponds to the fact that
some of the condensates have a net charge. 
Our two-species calculations show that, as one would expect,
when the pairing species have charges that do {\em not} cancel each other
there is a non-zero Meissner mass
for the gauge boson. When the pairing becomes gapless,
we find that this Meissner mass becomes imaginary. This confirms
existing \nova{zero-temperature} calculations for the full 
two-flavor three-color
(2SC/g2SC) and three-flavor three-color (CFL/gCFL) pairing patterns
\cite{Huang:2004am,Casalbuoni:2004tb,Giannakis:2004pf}.
\nova{We find that even in the gapped case, turning on a 
temperature in the appropriate range can cause the Meissner mass to
change from real to imaginary.}

\section{Two-species pairing formalism}
\label{sec:quasiparticles}

In our analysis, we will treat two massless
species of fermions (we will refer to them as ``quarks'')
that pair to yield a condensate
\beq
\langle \psi_a C\gamma_5 \psi_b \rangle = \De \eps_{ab} \ .
\label{condensate}
\eeq
By comparison with \Eqn{gCFL_condensate} and table~\ref{tab:charges}
we can see that the $rd$-$gu$, $ru$-$bs$,
and $gs$-$bd$ sectors of the CFL or gCFL quark condensate
have this structure.

Note that the Dirac charge conjugation matrix $C$
does not connect left-handed with right-handed quarks: the
pairing pattern is $\<\psi_L\psi_L\>$ and $\<\psi_R\psi_R\>$.
Similarly the gauge interactions preserve chiral symmetry,
and do not couple left-handed quarks to right-handed quarks.
A fermion mass term would couple left-handed to right-handed,
but most existing treatments of the gCFL phase work to
lowest order in the strange quark mass $M_s$, including it via
a chemical potential for strangeness 
$\dmu_s = - M_s^2/\mubar$,
which also does not couple left-handed quarks to right-handed quarks.
To this level of approximation, then, we can treat the left-handed
and right-handed quarks as completely decoupled from each other.
We can therefore reduce the 4-dimensional Dirac space to a
2-dimensional Weyl space.
To treat the quark-quark condensation, which violates
fermion number and allows quarks to turn into antiquarks,
we need to use Nambu-Gor'kov spinors, which
incorporate particles and antiparticles into the same spinor,
which doubles the size of our space. So we will work with
8-dimensional chiral spinors $\chi$, which arise from a tensor
product of the 2-dimensional Weyl space, the 2-dimensional flavor space,
and the Nambu-Gor'kov doubling.

Ignoring electromagnetism, the Lagrangian for our quarks is
\begin{equation}
{\cal L}_0=\frac{1}{2}\int \chi(p)^\dagger S^{-1}(p) \chi(p)\,dp \ ,
\end{equation}
where the inverse propagator is
\beq
S^{-1}(p)= 
\left(\ba{c@{\hspace{-2em}}c}
(p^0+\bar\mu)\,\one\otimes \one + {\vec p}\cdot{\vec \sigma}\otimes \one 
 +\delta \mu \, \one\otimes \sigma_3  & i\Delta\,\sigma_2\otimes \sigma_1 \\
-i\Delta\,\sigma_2^T\otimes \sigma_1 & (p^0-\bar\mu)\,\one\otimes \one 
 + {\vec p}\cdot{\vec \sigma^T}\otimes \one -\delta \mu \one\otimes \sigma_3
\ea\right)
\label{fermion_matrix}
\eeq
The Nambu-Gor'kov space is shown explicitly in the $2\times 2$ structure
of the matrix. In each entry, the first factor in the tensor product
lives in the 2-dimensional Weyl (spin) space, and the second factor 
lives in the 2-dimensional flavor space.
The only parameters are the average
chemical potential $\bar\mu$, the chemical potential splitting $\dmu$,
and the pairing gap parameter $\De$.
The off-diagonal terms
correspond to the quark pairing: they are proportional to $\De$,
with a Weyl factor of $\si_2$ because
the Dirac matrix $C= \diag(\si_2,\si_2)$ in the chiral basis,
and a flavor factor of $\si_1$, indicating that the
two flavors pair with each other, not with themselves.
The on-diagonal terms are standard free fermion terms.

The eight eigenvalues of $S^{-1}$ are
\begin{equation}
p^0 \pm \sqrt{(|{\vec p}| \pm \bar\mu)^2 + \Delta^2} \pm \delta \mu.
\end{equation}
The dispersion relations of the quasiquarks are given by
the poles in the propagator, i.e.~the zeros of $\det(S^{-1})$.
With the usual convention that negative energy states are filled, so that
all excitations have positive energy, we find
\begin{equation}
E(p) = \left| \sqrt{(|{\vec p}| \pm \bar\mu)^2 + \Delta^2} 
  \pm \delta \mu \right| \ .
\end{equation}
We see immediately how the relative sizes of $\De$ and $\dmu$
determine the form of the quasiparticle spectrum,
\beq
\ba{l@{\qquad}l@{\qquad}l}
\De > \dmu & \mbox{Gapped spectrum} 
  & \mbox{($rd$/$gu$ quarks in gCFL)} \\
\De = \dmu & \mbox{Ungapped quadratic spectrum} 
  & \mbox{($bu$/$rs$ quarks in gCFL)} \\
\De < \dmu & \mbox{Ungapped linear spectrum} 
  & \mbox{($gs$/$bd$ quarks in gCFL)} \\
\ea
\eeq
Actually, in gCFL the $bu$/$rs$ is not precisely gapless, but it
is very close: $\dmu$ is slightly bigger than $\De$,
\beq
0 < \de\mu-\De \ll \de\mu,\De \ll \bar\mu \ ,
\eeq
and we will study how the Debye and Meissner masses vary as the temperature
is scanned across the full range from $T\ll \dmu-\De$ to $T\gg\mubar$.
Note, however, that because we do not 
include the physics that leads to the Cooper pairing, $\De$ is just
a numerical parameter and it does not have the correct
$T$-dependence that would send it to zero at $T_c\approx 0.57\De$.
This means that our treatment of pairing is
only valid for $T\ll\De$ (which
is in any case the relevant regime for quark matter in neutron stars)
so insofar as our results depend on $\De$, they are only valid for
$T\ll\De$.

\section{Gauge boson self-energy}
\label{sec:self-energy}

We now discuss the in-medium properties of a gauge boson. These
depend on the charges of the two quark species. In general
their charges could be 
$(q_1,q_2)=\bar q \pm \de q$, and it turns out that
these contributions decouple from one another, so that for both
the Debye and Meissner masses we find
\beq
M^2 = {\bar q}^2 M_{\rm charged}^2   + (\de q)^2 M_{\rm neutral}^2
\eeq
Where $M_{\rm charged}$ is the result for a charged condensate in which
both flavors have the same charge, and $M_{\rm neutral}$ is the result for 
a neutral condensate in which
the two flavors have opposite charges.
Because of this decoupling, we will consider separately the case of a
neutral condensate with $(q_1,q_2)=(1,-1)$ and a charged condensate
with $(q_1,q_2)=(1,1)$.

\begin{figure}[tb]
\begin{center}
\includegraphics[width=0.5\textwidth]{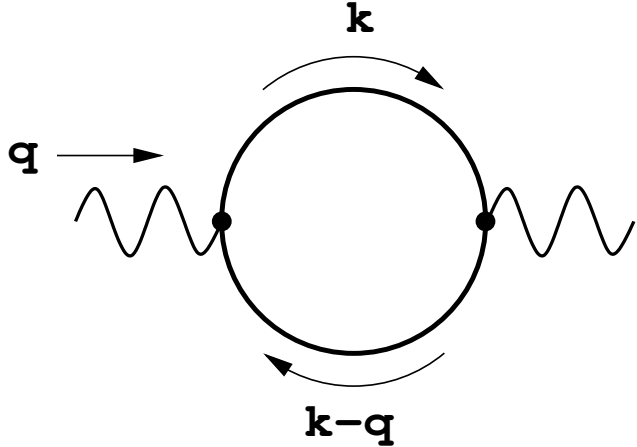}
\end{center}
\caption{\small
The Feynman diagram for the lowest-order contribution to the
photon self-energy. The internal lines are fermion propagators,
obtained by inverting \eqn{fermion_matrix}.
}
\label{fig:diagram}
\end{figure}

In the Nambu-Gor'kov formalism, the covariant coupling of the fermions
to the gauge boson takes the form
\beq
\frac{1}{2}\,e\, \chi^T \Ga^{\mu} \chi
\eeq
where the gauge coupling is $e$ and
$\Ga^\mu$ depends on the charges of the fermions:
\beq
\ba{rcl}
 \Ga^{\mu}_{\rm neutral} &=& \left( \ba{cc}   \si^{\mu}\otimes\si_3&0\\
                          0&-{\si^{\mu}}^T\otimes\si_3
            \ea\right), \\[2ex]
 \Ga^{\mu}_{\rm charged} &=& \left( \ba{cc}   \si^{\mu}\otimes\one&0\\
                          0&-{\si^{\mu}}^T\otimes \one
            \ea\right), \\
\si^{\mu} &=& \left(\one,{\vec \si}\right).
\ea
\label{vertices}
\eeq
To lowest order in the gauge coupling,
the gauge boson self-energy at external momentum $q$ is
\beq
\Pi_{\rm bare}^{\mu\nu}(q) = e^2 \int \frac{d^4k}{(2\pi)^4} 
  {\rm Tr}[\Ga^{\mu}S(k)\Ga^{\nu}S(k\!-\!q)].
\label{self-energy}
\eeq
This corresponds to the Feynman diagram shown in Fig.~\ref{fig:diagram}.
It is to be evaluated using the fermion propagators $S$ obtained by
inverting \eqn{fermion_matrix}. 
The technical details are given in the appendix.
Note that our expression \eqn{self-energy} lacks a factor
of $\half$ when compared with Eq.~(20) of Ref.~\cite{Rischke:2000qz}
or Eq.~(35) of Ref.~\cite{Huang:2004am}. This is because
we only include one chirality of fermion in our formalism, so
we multiply our result by 2 to obtain the value of $\Pi_{\rm bare}$
for a Dirac fermion with both chiralities.

For the transport properties of gCFL quark matter, we are interested
in low momentum properties, corresponding to the limit $q \to 0$.
For applications to neutron stars we are interested
in low temperatures $T\to 0$. We also expect singular behavior as
the quark spectrum goes quadratically gapless ($\dmu\to\De$)
as seen in the $bu$/$rs$ modes in gCFL. We have to be careful about the
ordering of all these limits.
We will take $q\to 0$ first, and then study a range of $T$, both
above and below  $|\dmu-\De|$.
Our calculations of $\Pi_{\rm bare}$ will yield divergent integrals
such as (\ref{Meissner_freqsum}).
To obtain physical results we require
a regularization and renormalization prescription. We use a
momentum cutoff, and subtract off the vacuum contribution, so that
our \nova{renormalized} results contain only the \nova{parts that depend on
$\mubar$, $\dmu$, and $T$},
\beq
\Pi(T,\bar\mu,\dmu,\De)  = \Pi_{\rm bare}(T,\bar\mu,\dmu,\De) 
- \Pi_{\rm bare}(0,0,0,\De)\ .
\label{subtraction}
\eeq
The renormalization subtraction term uses the same gap parameter $\De$
as the bare contribution.  Ultimately, the justification for this is
that we then get the correct value for the Meissner mass in the
presence of a neutral condensate, namely zero.
Other calculations in the literature \cite{Rischke:2000qz,Huang:2004am}
use a different prescription,
$\Pi(T,\bar\mu,\dmu,\De)  = \Pi_{\rm bare}(T,\bar\mu,\dmu,\De) 
- \Pi_{\rm bare}(0,0,0,0)$, which in our calculation
would yield a spurious contribution
of order $e^2\De^2$ to the Meissner mass. Such a contribution
can be seen in Rischke's result for the $\la_3$ gluon mass
in two-flavor quark matter (Eq.~(115) of Ref.~\cite{Rischke:2000qz}).
Rischke guessed that this contribution would be cancelled by other 
contributions that had been neglected in his calculation. However
we do not make any approximations (unlike Rischke we do not try to
combine a realistic pairing mechanism with our gauge boson mass
calculation, so our computations are simpler than his) 
and we can see that this term is not cancelled.
We think that our prescription is the correct one for
the situation that we study, but we do not claim
to have provided an {\em a priori} justification for it.

\section{Debye and Meissner mass for a neutral condensate}
\label{sec:neutral}

\subsection{Results}
\nova{The Debye and Meissner masses are defined by the 
static long-distance 
limit of the self energy $\Pi^{\mu\nu}(q)$. (Analysis of the self energy
for non-zero $q$ yields interesting information about how the photons
resolve the Cooper pairs \cite{Litim:2001mv}, but we do not
attempt such an analysis here.)
}
Taking $q\to 0$ in equations \eqn{Debye_freqsum} and \eqn{Meissner_freqsum}
and using \eqn{subtraction}, 
we obtain
\beq
\lim_{q\to 0}\Pi^{00}(q) = \frac{e^2}{\pi^2} \int_0^{\infty} dk \, k^2 
\Biggl\{\frac{d}{dE_+}\left[n_+(E_+)+ n_-(E_+)\right] 
   + \frac{d}{dE_-}\left[n_+(E_-)+ n_-(E_-)\right]\Biggr\} ,
\label{Debye_integral}
\eeq
and
\beq
\ba{rcl}
\dsp\lim_{q\to 0}\Pi^{ij}(q) &=&\dsp 2e^2 \int\frac{d^3k}{(2\pi)^3} 
  \Biggl\{ \hat{k}^i\hat{k}^j\left[\De^2\left(\frac{n_+(E_+) + n_-(E_+)}{E_+^3} 
       + \frac{n_+(E_-)+ n_-(E_-)}{E_-^3}\right)\right.\\
&&\left.\quad\dsp +\frac{(k+\bar\mu)^2}{E_+^2}\frac{d}{dE_+}\left[n_+(E_+)+ n_-(E_+)\right] + \frac{(k-\bar\mu)^2}{E_-^2}\frac{d}{dE_-}\left[n_+(E_-)+ n_-(E_-)\right]\right] \\
&&\dsp +(\de^{ij} -\hat{k}^i\hat{k}^j)\left[ \left(1-\frac{k^2-\bar\mu^2+\De^2}{E_+E_-}\right)\frac{n_+(E_+)+ n_-(E_+)-n_+(E_-)- n_-(E_-)}{E_+-E_-}\right.\\
&&\dsp \left.\quad+ \left(1+\frac{k^2-\bar\mu^2+\De^2}{E_+E_-}\right)\frac{n_+(E_+)+ n_-(E_+)+n_+(E_-)+ n_-(E_-)}{E_++E_-}\right]\Biggr\} \ ,
\label{Meissner_integral}
\ea
\eeq
where
\bea
E_{\pm}(k) &\equiv& \sqrt{(k\pm \bar\mu)^2+\De^2}\ ,\\
n_{\pm}(E) &\equiv& \dsp
  \left(\exp\Bigl(\frac{E\mp\dmu}{T}\Bigr)+1\right)^{-1}\ .
\eea
%


The integral (\ref{Meissner_integral}) evaluates to zero as one would expect:
a neutral condensate does not break gauge symmetries, so the Meissner mass
is zero,
\beq
M_M^2 = \frac{1}{2}\lim_{q\to 0}\left(\de_{ij}-\hat{q}_i\hat{q}_j\right)\Pi^{ij}(q) = 0 \ .
\label{neutral_Meissner}
\eeq
Note that 
in obtaining this result it was crucial that we used the correct 
renormalization subtraction \eqn{subtraction}.

The integral in (\ref{Debye_integral}) can be evaluated numerically,
and also analytically
in certain limits. For the Debye mass $M_D$, 
defined by $M_D^2=-\lim_{q\to 0}\Pi^{00}(q)$, we find
\beq
M_D^2 = \frac{e^2\bar\mu^2}{\pi^2}
\left\{
\begin{array}{ll@{\quad}l}
\dsp 2\left(1+\frac{\pi^2T^2}{3\bar\mu^2}\right), 
    & \dmu, \De \ll T,\bar\mu  & \mbox{(a)}\\[2ex]
\dsp \zeta\sqrt{\frac{2\De}{T}}, 
    & |\dmu-\De| \ll T \ll \De,\dmu \ll \bar\mu & \mbox{(b)}\\[2ex]
\dsp 2\left(1+\frac{\dmu^2\!-\!\De^2}{\bar\mu^2}\right) 
  \frac{\dmu}{\sqrt{\dmu^2\!-\!\De^2}}, 
    & T \ll |\dmu-\De|,\ \De < \dmu \ll \bar\mu & \mbox{(c)}\\[2ex]
\dsp \sqrt{\frac{2 \pi \De}{T}} \e^{-\frac{\De-\dmu}{T}}, 
    & T \ll |\dmu-\De|,\ \dmu <\De \ll \bar\mu  & \mbox{(d)}
\end{array}
\right.
\label{neutral_Debye}
\eeq
where the numerical constant $\zeta$ is
\beq
\zeta = \int_{-\infty}^{\infty} dx 
  \frac{\e^{-x^2}}{\left(\e^{-x^2}+1\right)^2} \approx 0.673718 \ .
\eeq
In the zero temperature limit this becomes
\beq
M_D^2(T=0) = \frac{2e^2}{\pi^2} (\bar\mu^2+\de\mu^2-\De^2) 
\frac{\de\mu}{\sqrt{\de\mu^2-\De^2}} \theta(\de\mu-\De) \ .
\label{neutral_Debye_T0}
\eeq

\subsection{Discussion}
\label{sec:discussion}

\begin{figure}[tb]
\begin{center}
\includegraphics[width=0.7\textwidth]{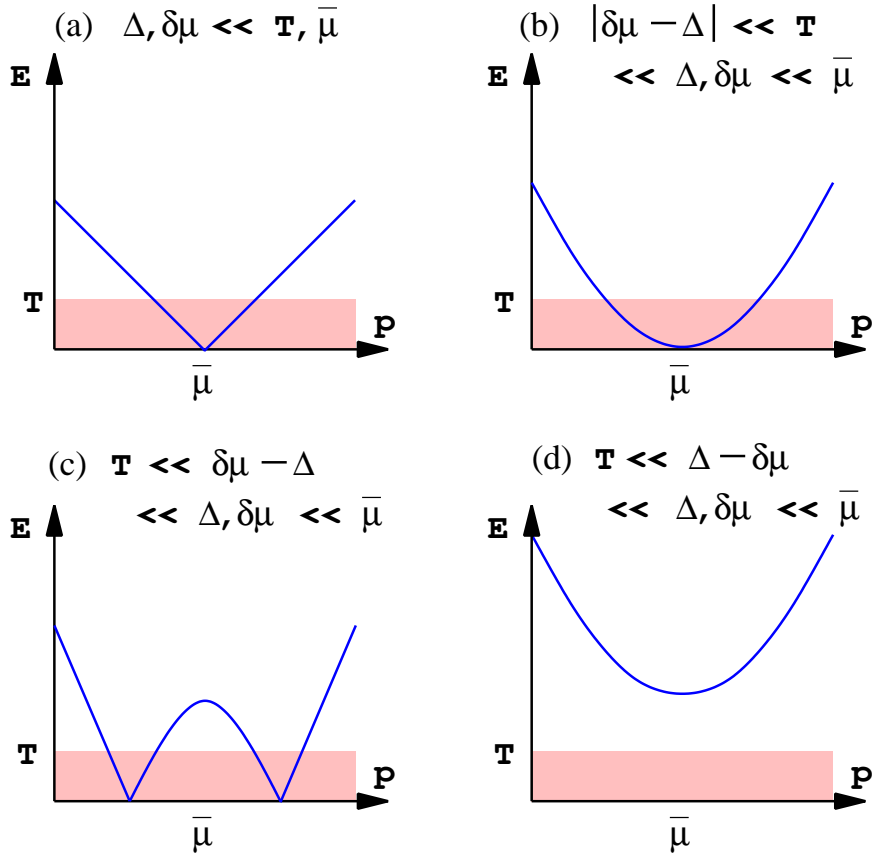}
\end{center}
\caption{\small
How the dispersion relations of the quasiparticles
look at energies of the same order as the temperature, for
temperatures coresponding to the cases distinguished in 
\Eqn{neutral_Debye}. The temperature decreases from (a) to (b)
to (c)/(d), and the energy scale stretches accordingly.
In each case the shaded region indicates energies $E < T$.
}
\label{fig:schematics}
\end{figure}

Our result for the Debye mass of a photon
passing through a neutral condensate of charged quarks
shows a subtle interplay of the
limits of small $T$ and small $|\dmu-\De|$. To understand why
the various limits behave so differently, it is useful to
recall how the dispersion relations look at energy of order
$T$ for the different ranges of temperature distinguished in
\Eqn{neutral_Debye}. These are shown in Fig.~\ref{fig:schematics}.
We will discuss each of the four regimes, bearing in mind that
the physical application of our result is to a
photon in gCFL quark matter,
in which the photon's in-medium properties
are dominated by the $bu$/$rs$ quarks 
with their near-quadratic gapless dispersion relation.

\begin{list}{} {
  \setlength{\topsep}{1ex} 
  \setlength{\itemsep}{-0.5\parsep} 
  \setlength{\labelwidth}{0em}
  \setlength{\labelsep}{0em}
  \setlength{\leftmargin}{0em} 
 }
\item[Case~(a)~] corresponds to high temperatures, where the structure
in the dispersion relations at scales of order $\De,\dmu$ is invisible,
so the system behaves like free particles with chemical potential
$\bar\mu$. As noted in section \ref{sec:self-energy}, our treatment of the
pairing is not valid in this range, but the result turns out to be independent
of pairing (i.e.~of $\De$). Because we have two species we get double the
standard one-species result, which is
$M_D^2 = e^2(\bar\mu^2/\pi^2+T^2/3)$ (Ref.~\cite{LeBellac}
Eq.~(6.103), in which $m_D^2=2m^2$, see Eq.~(7.125)).

\item[Case~(b)~] corresponds to temperatures much less than
$\dmu$ and $\De$, but much greater than their difference.
This is the typical situation for the $bu$/$rs$
quasiparticles of gCFL matter in a neutron star that is not very old.
At energies of the order of $T$ the dispersion relation looks
as if it barely touches the momentum axis at a double zero
(compare the $bu$/$rs$ line in Fig.~\ref{fig:disprel}).
This gives the unusual behavior $M_D^2 \propto 1/\sqrt{T}$:
the Debye mass {\em increases} as $T$ decreases.
\item[Case~(c)~] is for temperatures far below the splitting,
in the case where $\dmu>\De$. This corresponds to the $bu$/$rs$
quasiparticles of gCFL quark matter
in a very old, cold neutron star.
Now a typical thermal fluctuation can resolve the apparent
double zero into two separate zeroes. The Debye mass levels out
at a constant value as it would for free particles. If
$\De \ll \dmu \ll \bar\mu$ then this constant value is the
same as in case (a).
\item[Case~(d)~] is at a similar temperature to case (c),
but for a system where $\dmu<\De$, so the dispersion
relation never actually drops to zero. Relative to a typical
thermal fluctuation the quasiparticle gap is a large energy barrier, 
and $M_D$ drops to zero very rapidly with decreasing $T$.
\end{list}

\begin{figure}[tb]
\begin{center}
\includegraphics[width=0.7\textwidth]{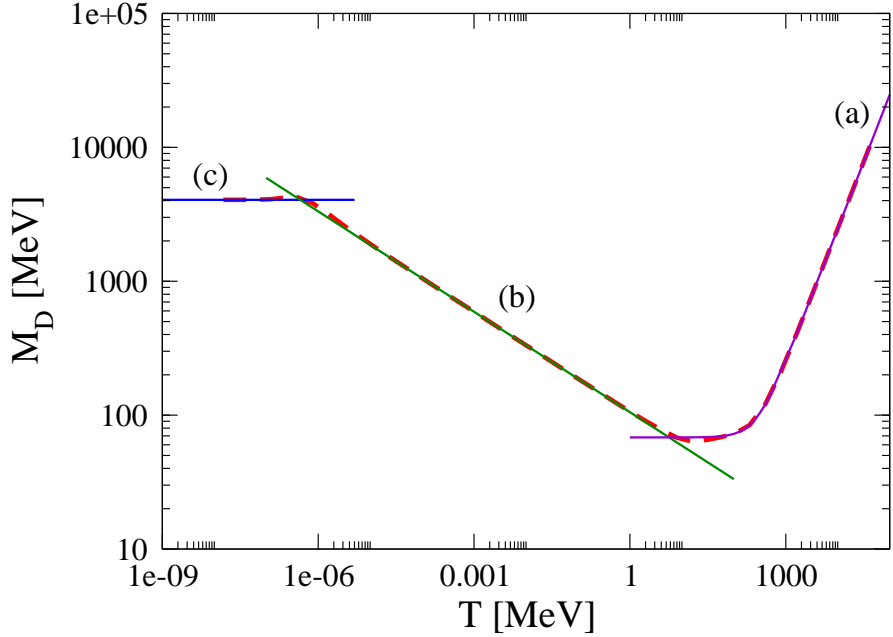}
\end{center}
\caption{\small
The Debye mass as a function of temperature,
for $\bar\mu=500~\MeV$, $\dmu\approx\De=25~\MeV$,
$\dmu-\De=10^{-6}~\MeV$. The heavy dashed line is the numerical
result \eqn{Debye_integral}. The solid lines show the 
analytic approximations in various regimes of Eq.~\eqn{neutral_Debye},
(a), (b), (c). The temperature of a neutron star varies from
$\sim 10~\MeV$ in the first seconds after the supernova
to $\sim \eV$ after hundreds of millions of years.
}
\label{fig:Debye_neutral}
\end{figure}

This gives us the temperature dependence for $M_D$ shown in
Fig.~\ref{fig:Debye_neutral}. We have chosen
values for the parameters that are appropriate for
gCFL matter in a neutron star: $e^2/(4\pi) = 1/137$,
$\bar\mu=500~\MeV$,  $\dmu\approx\De=25~\MeV$,
$\dmu-\De=10^{-6}~\MeV$ \cite{gCFL}.
We evaluate the integral
\eqn{Debye_integral} numerically, and compare with
the analytic approximations of \eqn{neutral_Debye}.
Note that $M_D$ reaches a very large value at $T=0$. This is
because at $T=0$, $M_D$ diverges  as
$\dmu/\sqrt{\dmu^2-\De^2}$ when $\De\to\dmu^-$ \eqn{neutral_Debye_T0}.

\begin{figure}[tb]
\begin{center}
\includegraphics[width=0.7\textwidth]{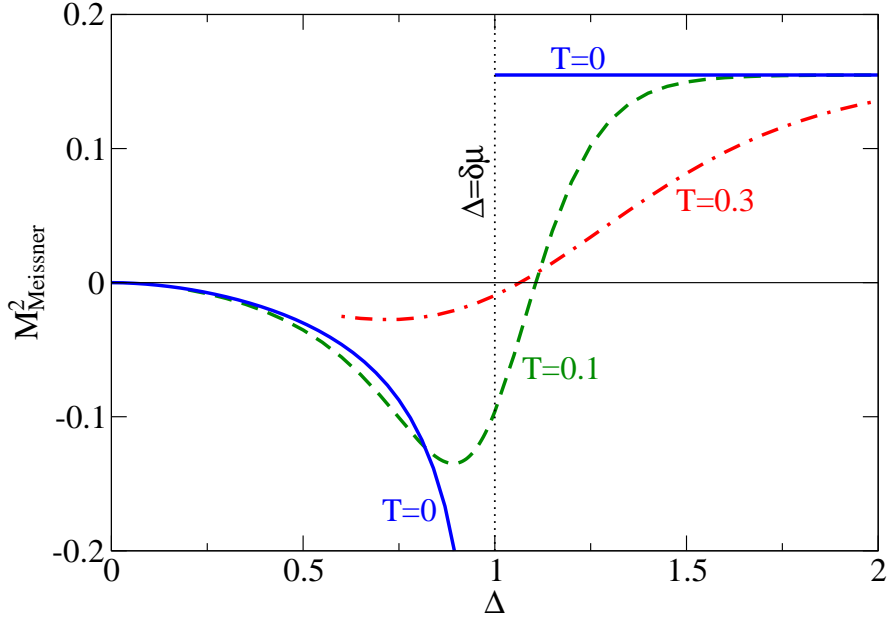}
\end{center}
\caption{\small
The square of the Meissner mass mass as a function of 
the gap parameter $\De$ at various temperatures, for 
$\bar\mu=5$, 
$\dmu=1$ in arbitrary units. 
\nova{Note how the zero-temperature discontinuity is smoothed
out at $T>0$. This means that in the gapped case ($\De>\dmu$)
turning on a temperature in the appropriate range
can cause the the Meissner mass 
to become imaginary.}
}
\label{fig:Meissner_Tcurves}
\end{figure}

\begin{figure}[tb]
\begin{center}
\includegraphics[width=0.7\textwidth]{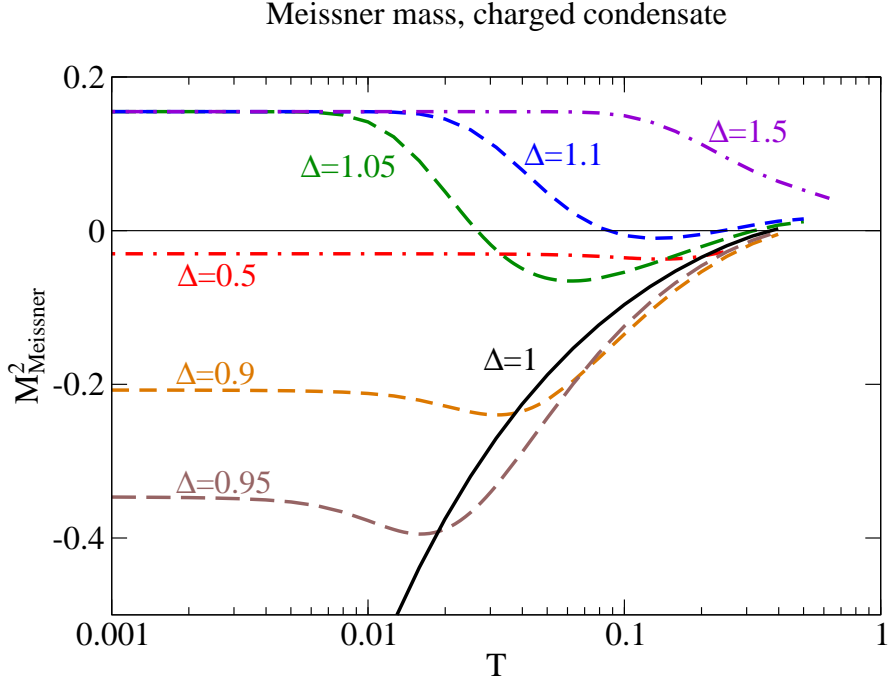}
\end{center}
\caption{\small
The square of the Meissner mass mass as a function of temperature
for various values of the gap parameter, with
$\bar\mu=5$, $\dmu=1$  in arbitrary units. 
\nova{As in Fig.~\ref{fig:Meissner_Tcurves}, we find that for the
gapped case ($\De>\dmu$), $M_M^2$ can go from a positive value at $T=0$
to a negative value when $T\gtrsim\De-\dmu$.
The curves stop when
$T$ reaches $0.5\De$, at which point the temperature-dependence of
$\De$ can no longer be neglected.}
}
\label{fig:Meissner_Decurves}
\end{figure}

\section{Debye and Meissner mass for a charged condensate}
\label{sec:charged}

\subsection{Results}

From \eqn{charged_Debye_freqsum} and \eqn{charged_Meissner_freqsum}, 
using \eqn{subtraction}, we get
\beq
\ba{rcl}
\dsp\lim_{q\to 0}\Pi^{00}(q) &=& \dsp -\frac{2e^2\bar\mu^2}{\pi^2}
  +\frac{e^2}{\pi^2} \int_0^{\infty} dk \, k^2 
  \Biggl\{\De^2\left[\frac{n_+(E_+)+n_-(E_+)}{E_+^3} 
  + \frac{n_+(E_-)+n_-(E_-)}{E_-^3}\right] \\
&&\dsp + \frac{(k+\bar\mu)^2}{E_+^2}\frac{d}{dE_+}\left[n_+(E_+)+ n_-(E_+)\right] 
  + \frac{(k-\bar\mu)^2}{E_-^2}\frac{d}{dE_-}\left[n_+(E_-)+ n_-(E_-)\right]
  \Biggr\} ,
\ea
\label{charged_Debye}
\eeq
and
\bea
\lim_{q\to 0}\Pi^{ij}(q) &=& \frac{2e^2\bar\mu^2\de^{ij}}{3 \pi^2} + 2e^2 \int \frac{d^3k}{(2\pi)^3}  
\Biggl\{\hat{k}^i\hat{k}^j \left[\frac{d}{dE_+}\left[n_+(E_+)+ n_-(E_+)\right] +\frac{d}{dE_-}\left[n_+(E_-)+ n_-(E_-)\right]\right]\nonumber\\
&&\quad +(\de^{ij} -\hat{k}^i\hat{k}^j)\left[ \left(1-\frac{k^2-\bar\mu^2-\De^2}{E_+E_-}\right)\frac{n_+(E_+)+ n_-(E_+)-n_+(E_-)- n_-(E_-)}{E_+-E_-}\right.\nonumber\\
&&\qquad \left.+ \left(1+\frac{k^2-\bar\mu^2-\De^2}{E_+E_-}\right)\frac{n_+(E_+)+ n_-(E_+)+n_+(E_-)+ n_-(E_-)}{E_++E_-}\right]\Biggr\}. 
\label{charged_Meissner}
\eea
At arbitrary temperature these integrals can be evaluated numerically.
In the zero temperature limit they reduce to
\beq
\ba{rcl}
M_D^2(T=0) &=& \dsp \frac{2e^2}{\pi^2}
 \Biggl\{ \bar\mu^2 + \theta(\de\mu-\De)\biggl(\de\mu\sqrt{\de\mu^2-\De^2} 
  + \De^2 \log\Bigl(\frac{\de\mu -\sqrt{\de\mu^2-\De^2}}{\De}\Bigr)\biggr)
    \Biggr\} \\[1ex]
M_M^2(T=0) &=& \dsp \frac{2e^2}{3\pi^2} 
  \Biggl\{\bar\mu^2 - \theta(\de\mu-\De)\biggl(\bar\mu^2\frac{\de\mu}{\sqrt{\de\mu^2-\De^2}} 
  -3 \De^2 \log\Bigl(\frac{\de\mu -\sqrt{\de\mu^2-\De^2}}{\De}\Bigr)\biggr)
    \Biggr\}.
\ea
\label{charged_T0}
\eeq
\nova{
When $\dmu<\De$ we recover the well-known result for gluons
in the CFL phase, $M_M^2/M_D^2 = 1/3$ \cite{Rischke:2000ra,Schmitt:2003aa}.
}

\subsection{Discussion}
For a given $\dmu$, we can calculate the Debye and Meissner masses as a 
function of temperature $T$ and the gap parameter $\De$. The Debye mass is
large, $M_D\gtrsim\bar\mu\gg \De$, so we do not discuss it in detail.
In the case of the Meissner mass, we can see that at $T=0$ we get an imaginary
value when $\dmu>\De$, i.e.~when the system is gapless. This is reminiscent of
results found by Shovkovy and Huang and others for 
gluons in the more complicated cases of g2SC and gCFL quark matter
\cite{Huang:2004am,Casalbuoni:2004tb,Giannakis:2004pf}.
The square of the Meissner mass as a function of
temperature is shown in 
Figs.~\ref{fig:Meissner_Tcurves} and \ref{fig:Meissner_Decurves}.
To make these plots we set $\bar\mu=5,\ \dmu=1$, in arbitrary energy units,
and used $e^2/(4\pi) = 1/137$ as in section \ref{sec:neutral}.
The solid ``$T=0$'' curve in Fig.~\ref{fig:Meissner_Tcurves} shows
the result of \Eqn{charged_T0}: the Meissner mass $M_M$ is zero when there
is no pairing ($\De=0$) \footnote{\nova{
Technically our renormalization condition \eqn{subtraction} corresponds
to throwing away the part of the $\De$-dependence that is independent
of $\mu,T,\dmu$. But this is always zero
because free fermions at $T=\mu=0$ have
no Meissner mass. So it is legitimate to
calculate the $\De$-dependence in our framework.
}}. When $\De<\dmu$, $M_M^2$ is negative, and
diverges to $-\infty$ as $\De\to\dmu^-$. At $\De=\dmu$,
$M_M^2$ jumps discontinuously to the positive value $2 e^2\bar\mu^2/(3\pi^2)$
that is characteristic of the spontaneous breaking of a gauged symmetry
by a charged condensate. It is clear that the limit $\De\to\dmu$
is singular, with very different behavior according to whether it
is taken from above or below. This is quite natural, since for
$\De>\dmu$ the system is always gapped, whereas for $\De\leqslant\dmu$
it is gapless. We expect that this singular behavior will be
smoothed out at any finite nonzero temperature, and this is
indeed the case. The curves for $T=0.1$ and $T=0.3$
in Fig.~\ref{fig:Meissner_Tcurves}
show a smooth transition, over a range $\de \De \sim T$, from
the gapless behavior to the gapped behavior.

Essentially the same information is presented in a different way
in Fig.~\ref{fig:Meissner_Decurves}, where we show the dependence on
$T$ for a range of different values of $\De$. Again, we fix
$\bar\mu=5,\ \dmu=1$. For $\De>\dmu$ (gapped system) we see $M_M^2$ tends to a
positive constant as $T\to 0$, whereas for $\De<\dmu$
we find that $M_M^2$ tends to a negative value that is large for $\De$
just below $\dmu$, but tends to zero as $\De\to 0$.
\nova{It is also apparent that the smoothing effect of the non-zero temperature
can change a positive $M_M^2$ into a negative one. This effect is also visible
in Fig.~\ref{fig:Meissner_Tcurves}, and is quite reasonable, given the
picture outlined in Fig.~\ref{fig:schematics}. If $\De>\dmu$ then the
dispersion relations are gapped, with positive $M_M^2$ at $T=0$;
but if $T>|\dmu-\De|$ then
to within the natural energy resolution
($T$, approximately) the dispersion relation looks quadratically gapless
and the Meissner mass will become imaginary.
}
It should be remembered that we have not included any of the
pairing dynamics in this calculation, so our gap parameters
are independent of temperature. 
\nova{The curves in Figs.~\ref{fig:Meissner_Tcurves}
and \ref{fig:Meissner_Decurves} stop when $T$ reaches $0.5\De$, 
at which point the temperature-dependence of
$\De$ can no longer be neglected.}

\section{Conclusions}
\label{sec:conclusions}

We have calculated the Debye and Meissner masses of a gauge
boson in the presence of a condensate of Cooper pairs that involve
two species of massless charged spin-$\half$ fermions.
We did not specify any particular pairing mechanism, but simply
parameterized the dispersion relations of the quasiparticles
using a momentum-independent gap parameter $\De$, and individual
chemical potentials $\mubar\pm\dmu$ for the two species.
We allowed the fermions to have arbitrary charge in their coupling to
the gauge boson, but found that this reduced to two elementary
cases: a neutral condensate (charges $(+1,-1)$), and a charged
condensate (charges $(+1,+1)$).
Our results for the neutral condensate are presented in section 
\ref{sec:neutral}
and our results for the charged condensate in section \ref{sec:charged}.

For the neutral condensate, our results give the in-medium behavior
of a very low energy
photon (technically, it is the massless $\Qt$ gauge boson that is
predominantly
the photon with a small admixture of gluon) in the gCFL phase of
quark matter. This is dominated by the gapless charged excitations,
the $bu$/$rs$ quasiquarks, which form a two-species system of the
type that we studied, where the strange quark mass has been treated
in lowest order as a contribution to the chemical potential for
strangeness. Because the photon coupling is weak, our calculation,
which includes only the leading order diagram 
(Fig.~\ref{fig:diagram}),
gives the dominant contribution.
We find that  the Debye mass
shows a non-monotonic temperature dependence, dropping to
a minimum value of order $\sqrt{\alpha}\mubar$ at temperatures 
\nova{somewhat below} 
the pairing gap. We find that the Meissner mass is zero,
as expected for a condensate that does not break the gauge symmetry,
and we note that to obtain this result it was necessary to use
the renormalization subtraction \eqn{subtraction}, which differs
from the one used in the existing literature on quark matter.

For the charged condensate, our results give some insight into the
in-medium behavior of gluons in color-superconducting phases.
The color-diagonal gluons can be treated
as photons with appropriate charge assignments to the various quarks
so that the condensates
in all three $2\times 2$ sectors of
the color-flavor-locked pairing pattern (Table~\ref{tab:charges})
carry net color charge, except that the $rd$-$gu$ condensate
is neutral to the $\la_3$ gluon.
We find that there is always a large Debye mass, of order $\mubar^2$ or
greater.
At zero temperature, the square of the Meissner mass is {\em negative} for
$0<\De\leqslant\dmu$, diverging to $-\infty$ as $\De\to\dmu^-$,
but then jumps discontinuously to a positive value for
$\De>\dmu$. The Meissner mass is therefore imaginary whenever the
quasiquark spectrum is gapless, and positive when the spectrum is
gapped. 
\nova{At $T>0$ this discontinuity is smoothed out,
with interesting consequences. For gapped systems
($\De>\dmu$) the Meissner mass is real at $T=0$, but it becomes
imaginary when $T\gtrsim\De-\dmu$, as the temperature-smeared dispersion
relation is then indistinguishable from a gapless one.}

Our charged-condensate results confirm the essential conclusion of
Refs.~\cite{Huang:2004am} and \cite{Casalbuoni:2004tb}, that
charged condensates with gapless excitations are associated with
imaginary Meissner mass. In particular, our results for the
Meissner mass agree well with those obtained for the diagonal
``$\tilde 8$ gluon'' (the combination of a
gluon and a photon that is orthogonal to $\Qt$)
in Ref.~\cite{Huang:2004am}.
It is interesting to note that
Ref.~\cite{Huang:2004am} also finds imaginary Meissner masses 
for the off-diagonal gluons in the g2SC phase,
even in situations where no quasiquark modes
are gapless. We cannot offer any insight into this because 
we only treat diagonal gauge bosons%
\footnote{\nova{
Our formalism can easily accomodate an off-diagonal coupling. We have
performed preliminary calculations for this case, but find an
unphysical $\dmu$-dependent logarithmic divergence.
This was present but discarded in the recent 2-flavor 
(g2SC) calculation \cite{Huang:2004am},
and is also thought to occur in the analogous gCFL calculation \cite{ASchmitt}.
}}.
Also, it is curious that Ref.~\cite{Casalbuoni:2004tb},
which treats the gCFL phase, only finds an 
imaginary Meissner mass for the off-diagonal ($\la_1$ and $\la_2$) gluons.
Although our 2-species calculation is not directly applicable to the
gCFL case, which has multiple pairing sectors (table \ref{tab:charges})
coupled together by neutrality constraints,
our results would lead us to expect that they should have found
imaginary Meissner masses for the
diagonal $\la_3$ and $\tilde 8$ gluons, which both
couple to the gapless $bu$-$rs$ and $gs$-$bd$ condensates.

Although we have couched our discussion in terms of gauge boson masses,
the quantity that we calculated, the 
low-momentum current-current two-point function, also has physical
meaning if the currents in question are not coupled to gauge fields.
In this case $M_M^2/(e^2\De^2)$ is the 
coefficient of the 
gradient term in the effective theory of small fluctuations around the
ground-state condensate.
The fact that we find a negative value when the quasiparticles are gapless
indicates an instability 
towards spontaneous breaking of translational invariance.
The nature of the true ground state in gapless color superconductors
remains unknown: it could be
a mixed phase \cite{Reddy:2004my} or a crystalline (LOFF) phase
\cite{Giannakis:2004pf}.
Since the two-species system
shows this instability, it may provide a convenient
toy model for investigating the nature of the true ground state.

\bc
{\bf Acknowledgements}
\ec
We are grateful to M.~Mannarelli for 
drawing our attention to the temperature-induced imaginary Meissner mass.
We have benefitted from discussions with M. Forbes, K.~Fukushima, D.~Hong,
K.~Rajagopal, D.~Rischke, A.~Schmitt, and I.~Shovkovy.
This research was supported by the Department of Energy
under grant number DE-FG02-91ER40628.

\appendix
\section{Calculational details}
\label{sec:appendix}

\subsection{Fermion propagator}
We calculate the fermion propagator by inverting $S^{-1}$ in \eqn{fermion_matrix}. To doing so, let us write $S$ as a $2\!\times\!2$ matrix
\beq
S=\left(
\ba{cc}
S_{11}&S_{12}\\
S_{21}&S_{22}
\ea
\right).
\eeq
Since $S$ is Hermitian, we have 
\beq
S_{12}=S_{21}^{\dag}.
\eeq
We may expand each matrix element in $S$ as 
\bea
S_{11}&=&\al'(p)\openone\otimes\openone + \be'(p)\openone\otimes\si_3 + \ga'(p) {\hat p}\cdot {\vec \si} \otimes\openone + \eta'(p) {\hat p}\cdot {\vec \si} \otimes \si_3;\nonumber\\
S_{12}&=&-i\De\left[a(p)\openone\otimes \openone+ b(p)\openone\otimes\si_3 + c(p) {\hat p}\cdot {\vec \si} \otimes\openone + d(p) {\hat p}\cdot {\vec \si} \otimes \si_3\right]\left(\si_2\otimes\si_1\right);\nonumber\\
S_{22}&=&\al(p)\openone\otimes\openone + \be(p)\openone\otimes\si_3 + \ga(p) {\hat p}\cdot {\vec \si^T} \otimes\openone + \eta(p) {\hat p}\cdot {\vec \si^T} \otimes \si_3.
\eea
After a lengthy calculation, we have
\bea
a(p) &=& \frac{1}{4D_1(p)} + \frac{1}{4D_2(p)} + \frac{1}{4D_3(p)}+ \frac{1}{4D_4(p)}\nonumber\\
b(p) &=& \frac{1}{4D_1(p)} + \frac{1}{4D_2(p)} - \frac{1}{4D_3(p)} - \frac{1}{4D_4(p)}\nonumber\\
c(p) &=& \frac{1}{4D_1(p)} - \frac{1}{4D_2(p)} + \frac{1}{4D_3(p)} - \frac{1}{4D_4(p)}\nonumber\\
d(p) &=& \frac{1}{4D_1(p)} - \frac{1}{4D_2(p)} - \frac{1}{4D_3(p)}+ \frac{1}{4D_4(p)},
\eea
\bea
\al(p) &=& \frac{p^0 + \dmu + |{\vec p}| + \bar\mu}{4 D_1(p)} + \frac{p^0 + \dmu - |{\vec p}| + \bar\mu}{4 D_2(p)} + \frac{p^0 - \dmu + |{\vec p}| + \bar\mu}{4 D_3(p)} + \frac{p^0 - \dmu - |{\vec p}| + \bar\mu}{4 D_4(p)} \nonumber\\
\be(p) &=& -\frac{p^0 + \dmu + |{\vec p}| + \bar\mu}{4 D_1(p)} - \frac{p^0 + \dmu - |{\vec p}| + \bar\mu}{4 D_2(p)} + \frac{p^0 - \dmu + |{\vec p}| + \bar\mu}{4 D_3(p)} + \frac{p^0 - \dmu - |{\vec p}| + \bar\mu}{4 D_4(p)} \nonumber\\
\ga(p) &=& -\frac{p^0 + \dmu + |{\vec p}| + \bar\mu}{4 D_1(p)} + \frac{p^0 + \dmu - |{\vec p}| + \bar\mu}{4 D_2(p)} - \frac{p^0 - \dmu + |{\vec p}| + \bar\mu}{4 D_3(p)} + \frac{p^0 - \dmu - |{\vec p}| + \bar\mu}{4 D_4(p)} \nonumber\\
\eta(p) &=& \frac{p^0 + \dmu + |{\vec p}| + \bar\mu}{4 D_1(p)} - \frac{p^0 + \dmu - |{\vec p}| + \bar\mu}{4 D_2(p)} - \frac{p^0 - \dmu + |{\vec p}| + \bar\mu}{4 D_3(p)} + \frac{p^0 - \dmu - |{\vec p}| + \bar\mu}{4 D_4(p)} ,\nonumber\\
\eea
and
\bea
\al'(p) &=& \frac{p^0 + \dmu - |{\vec p}| - \bar\mu}{4 D_1(p)} + \frac{p^0 + \dmu + |{\vec p}| - \bar\mu}{4 D_2(p)} + \frac{p^0 - \dmu - |{\vec p}| - \bar\mu}{4 D_3(p)} + \frac{p^0 - \dmu + |{\vec p}| - \bar\mu}{4 D_4(p)} \nonumber\\
\be'(p) &=&  \frac{p^0 + \dmu - |{\vec p}| - \bar\mu}{4 D_1(p)} + \frac{p^0 + \dmu + |{\vec p}| - \bar\mu}{4 D_2(p)} - \frac{p^0 - \dmu - |{\vec p}| - \bar\mu}{4 D_3(p)} - \frac{p^0 - \dmu + |{\vec p}| - \bar\mu}{4 D_4(p)} \nonumber\\
\ga'(p) &=&  \frac{p^0 + \dmu - |{\vec p}| - \bar\mu}{4 D_1(p)} - \frac{p^0 + \dmu + |{\vec p}| - \bar\mu}{4 D_2(p)} + \frac{p^0 - \dmu - |{\vec p}| - \bar\mu}{4 D_3(p)} - \frac{p^0 - \dmu + |{\vec p}| - \bar\mu}{4 D_4(p)} \nonumber\\
\eta'(p) &=&  \frac{p^0 + \dmu - |{\vec p}| - \bar\mu}{4 D_1(p)} - \frac{p^0 + \dmu + |{\vec p}| - \bar\mu}{4 D_2(p)} - \frac{p^0 - \dmu - |{\vec p}| - \bar\mu}{4 D_3(p)} + \frac{p^0 - \dmu + |{\vec p}| - \bar\mu}{4 D_4(p)},\nonumber\\
\eea
where
\bea
D_1(p) &=& (p^0+\dmu)^2-(|{\vec p}|+\bar\mu)^2-\De^2\nonumber\\
D_2(p) &=& (p^0+\dmu)^2-(|{\vec p}|-\bar\mu)^2-\De^2\nonumber\\
D_3(p) &=& (p^0-\dmu)^2-(|{\vec p}|+\bar\mu)^2-\De^2\nonumber\\
D_4(p) &=& (p^0-\dmu)^2-(|{\vec p}|-\bar\mu)^2-\De^2.
\eea

\subsection{Matsubara frequency sums}
To compute $\Pi_{\rm bare}^{\mu\nu}$, we need do evaluate the 4-dimensional 
momentum-space integral.
At finite temperature, the $k_0$ integral
becomes a Matsubara frequency sum,
in which we replace $k^0$ by $i\om_n$, and $\int d k^0$ by
$2\pi T\sum_{\om_n}$. Because quarks are fermions, they obey antiperiodic
temporal boundary conditions, and the Matsubara
frequencies are

\beq
\om_n=\frac{1}{T}(2 n+1)\pi, \qquad n \in \mathbb{Z}.
\eeq 
To evaluate the frequency sums, we use the following result
(Eq.\ (5.77) in Ref.~\cite{LeBellac}):
\bea
&&T\sum_{\om_n} \frac{1}{(i \om_n + \dmu)^2 + E_1^2}\, \frac{1}{(i \om_n- i\om + \dmu)^2 +E_2^2}\nonumber\\
&=& \frac{1}{4E_1E_2}\left[\frac{f_-(E_1)-f_-(E_2)}{i \om + E_1-E_2} - \frac{f_+(E_1)-f_+(E_2)}{i \om - E_1+E_2}\right.\nonumber\\
&&\qquad \left.+\frac{1-f_-(E_1)-f_+(E_2)}{i \om + E_1+E_2} - \frac{1-f_+(E_1)-f_-(E_2)}{i \om - E_1-E_2}\right],
\label{freqsum_quadratic}
\eea
where with positive $E$, the functions $f_\pm$ are defined as
\beq
f_{\pm}(sE) \equiv \frac{1}{\e^{\frac{sE\mp\dmu}{T}}+1}, \qquad s=\pm1.
\eeq
From Eq.\ (5.76) in Ref.~\cite{LeBellac}, we obtain another useful formula 
\beq
T\sum_{\om_n} \frac{1}{i \om_n + \dmu +s_1 E_1} \frac{1}{i (\om_n-\om) + \dmu +s_2 E_2}
= \frac{f_-(s_1 E_1)-f_-(s_2 E_2)}{i \om + s_1 E_1-s_2 E_2}, \quad s_1, s_2 = \pm1.
\label{freqsum_linear}
\eeq 
To apply these formulas directly, we need to use the
method of partial fractions to convert expressions with
$k^0$ in the numerator to expressions with $k^0$ appearing only
in the denominator. A typical example is
\bea
&&\frac{(k^0+\dmu)(k^0-q^0+\dmu)}{D_1(k)D_1(k\!-\!q)}\nonumber\\
&=& \frac{1}{4D_1(k)D_1(k\!-\!q)} \left\{\left[k^0 + \dmu + \sqrt{(|\vec{k}|+\bar\mu)^2 + \De^2}\right] + \left[k^0+ \dmu - \sqrt{(|\vec{k}|+\bar\mu)^2 + \De^2}\right]\right\}\nonumber\\
&&\quad \times \left\{\left[k^0 -q^0 + \dmu + \sqrt{(|\vec{k}-\vec{q}|+\bar\mu)^2 + \De^2}\right] + \left[k^0-q^0+ \dmu - \sqrt{(|\vec{k}-\vec{q}|+\bar\mu)^2 + \De^2}\right]\right\}\nonumber\\
&=& \frac{1}{4} \Biggl[\frac{1}{k^0+ \dmu + \sqrt{(|\vec{k}|+\bar\mu)^2 + \De^2}} \frac{1}{k^0 -q^0 + \dmu + \sqrt{(|\vec{k}-\vec{q}|+\bar\mu)^2 + \De^2}}   \nonumber\\
&& \quad + \frac{1}{k^0+ \dmu + \sqrt{(|\vec{k}|+\bar\mu)^2 + \De^2}} \frac{1}{k^0 -q^0 + \dmu - \sqrt{(|\vec{k}-\vec{q}|+\bar\mu)^2 + \De^2}}   \nonumber\\
&& \quad +\frac{1}{k^0+ \dmu - \sqrt{(|\vec{k}|+\bar\mu)^2 + \De^2}} \frac{1}{k^0 -q^0 + \dmu + \sqrt{(|\vec{k}-\vec{q}|+\bar\mu)^2 + \De^2}}   \nonumber\\
&& \quad  + \frac{1}{k^0+ \dmu - \sqrt{(|\vec{k}|+\bar\mu)^2 + \De^2}} \frac{1}{k^0 -q^0 + \dmu - \sqrt{(|\vec{k}-\vec{q}|+\bar\mu)^2 + \De^2}} \Biggr].
\eea
With the integrand in this form, the frequency sum can be evaluated.

\subsection{Neutral condensate}
We substitute the fermion propagator into the expression for the photon 
self-energy \eqn{self-energy} with $\Ga^{\mu}_{\rm neutral}$ in \eqn{vertices}. The zero-zero component is
\bea
\Pi_{\rm bare}^{00} 
&=& e^2 \int \frac{d^4k}{(2\pi)^4} \Biggl\{\left[1+{\hat k}\cdot(\widehat{k\!-\!q})\right] 
\frac{(k^0+\dmu)(k^0-q^0+\dmu)+\left(|{\vec k}|+\bar\mu\right)\left(|{\vec k}-{\vec q}|+\bar\mu\right) +\De^2}{D_1(k)D_1(k\!-\!q)}\nonumber\\
&& \qquad +\left[1-{\hat k}\cdot(\widehat{k\!-\!q})\right] 
\frac{(k^0+\dmu)(k^0-q^0+\dmu)-\left(|{\vec k}|+\bar\mu\right)\left(|{\vec k}-{\vec q}|-\bar\mu\right) +\De^2}{D_1(k)D_2(k\!-\!q)}\nonumber\\
&& \qquad + (\dmu\to\dmu)(\bar\mu\to-\bar\mu) + (\dmu\to-\dmu)(\bar\mu\to\bar\mu) + (\dmu\to-\dmu)(\bar\mu\to-\bar\mu)\Biggr\},
\eea
where $(\dmu\to\pm\dmu)(\bar\mu\to\pm\bar\mu)$ means terms as the first two terms with $\dmu$ replaced by $\pm\dmu$ and $\bar\mu$ replaced by $\pm\bar\mu$. 

The spatial components of the photon self-energy are
\bea
 \Pi_{\rm bare}^{ij}&=& e^2 \int \frac{d^4k}{(2\pi)^4} \Biggl\{\nonumber\\
&& \quad \de^{ij} \left[\left(1-{\hat k}\cdot(\widehat{k\!-\!q})\right) 
\frac{(k^0+\dmu)(k^0-q^0+\dmu)+\left(|{\vec k}|+\bar\mu\right)\left(|{\vec k}-{\vec q}|+\bar\mu\right) -\De^2}{D_1(k)D_1(k\!-\!q)}\right.\nonumber\\
&& \qquad \left.+\left(1+{\hat k}\cdot(\widehat{k\!-\!q})\right)
\frac{(k^0+\dmu)(k^0-q^0+\dmu)-\left(|{\vec k}|+\bar\mu\right)\left(|{\vec k}-{\vec q}|-\bar\mu\right) -\De^2}{D_1(k)D_2(k\!-\!q)}\right]\nonumber\\
&& \quad +\left[{\hat k}^i(\widehat{k\!-\!q})^j + (\widehat{k\!-\!q})^i{\hat k}^j\right]\left[\frac{(k^0+\dmu)(k^0-q^0+\dmu)+\left(|{\vec k}|+\bar\mu\right)\left(|{\vec k}-{\vec q}|+\bar\mu\right) -\De^2}{D_1(k)D_1(k\!-\!q)}\right.\nonumber\\
&& \qquad \left.-\frac{(k^0+\dmu)(k^0-q^0+\dmu)-\left(|{\vec k}|+\bar\mu\right)\left(|{\vec k}-{\vec q}|-\bar\mu\right) -\De^2}{D_1(k)D_2(k\!-\!q)}\right]\nonumber\\
&& \quad +(\bar\mu\to-\bar\mu)\nonumber\\
&& \quad + i \eps^{ijl}\Biggl[\left({\hat k}_l-(\widehat {k\!-\!q})_l\right)\left(-\frac{(k^0+\dmu)(k^0-q^0+\dmu)+\left(|{\vec k}|+\bar\mu\right)\left(|{\vec k}-{\vec q}|+\bar\mu\right) -\De^2}{D_1(k)D_1(k\!-\!q)}\right.\nonumber\\
&& \qquad \quad \left. + \frac{(k^0+\dmu)(k^0-q^0+\dmu)+\left(|{\vec k}|-\bar\mu\right)\left(|{\vec k}-{\vec q}|-\bar\mu\right) -\De^2}{D_2(k)D_2(k\!-\!q)}\right)\nonumber\\
&&\qquad +\left({\hat k}_l+(\widehat {k\!-\!q})_l\right)\left(-\frac{(k^0+\dmu)(k^0-q^0+\dmu)-\left(|{\vec k}|+\bar\mu\right)\left(|{\vec k}-{\vec q}|+\bar\mu\right) -\De^2}{D_1(k)D_2(k\!-\!q)}\right.\nonumber\\
&& \qquad \quad\left. + \frac{(k^0+\dmu)(k^0-q^0+\dmu)-\left(|{\vec k}|-\bar\mu\right)\left(|{\vec k}-{\vec q}|-\bar\mu\right) -\De^2}{D_2(k)D_1(k\!-\!q)}\right)\Biggr]\nonumber\\
&& \quad +(\dmu\to-\dmu)\Biggr\},
\eea
where $(\bar\mu\to-\bar\mu)$ and $(\dmu\to-\dmu)$ means a duplication of
{\em all} previous terms with $\bar\mu$ replaced by $-\bar\mu$ 
or $\dmu$ replaced by $-\dmu$.

Performing the frequency sums using \eqn{freqsum_quadratic} and \eqn{freqsum_linear}, we obtain the zero-zero component of the photon self-energy:
\beq
\ba{rl}
\dsp \Pi_{\rm bare}^{00} = \frac{e^2}{4} \int \frac{d^3k}{(2\pi)^3} \Bigl\{
& \left[1+{\hat k}\cdot(\widehat{k\!-\!q})\right]
  \left[1+f_{+,+}^{(1)}(k,p)\right]u_{+,+}(k,p)\\[1ex]
 + & \left[1+{\hat k}\cdot(\widehat{k\!-\!q})\right]
   \left[1-f_{+,+}^{(1)}(k,p)\right]v_{+,+}(k,p)\\[1ex]
 + & \left[1-{\hat k}\cdot(\widehat{k\!-\!q})\right] 
   \left[1+f_{+,-}^{(1)}(k,p)\right]u_{+,-}(k,p)\\[1ex]
 + & \left[1-{\hat k}\cdot(\widehat{k\!-\!q})\right]
   \left[1-f_{+,-}^{(1)}(k,p)\right]v_{+,-}(k,p)\\
 + &  (\de\mu\to\de\mu)(\bar\mu\to-\bar\mu) + (\de\mu\to-\de\mu)(\bar\mu\to\bar\mu) \\
 + &  (\de\mu\to-\de\mu)(\bar\mu\to-\bar\mu)\Bigr\},
\ea
\label{Debye_freqsum}
\eeq
where
\beq
\ba{rcl}
 f_{r,s}^{(1)}(k,q) &\equiv& \dsp  \frac{\left(r|{\vec 
k}|+\bar\mu\right)\left(s|{\vec k}-{\vec 
q}|+\bar\mu\right)+\De^2}{E_r(k)E_s(k\!-\!q)}, \\[2ex]
u_{r,s}(k,q) &\equiv& \dsp \frac{n_-[E_r(k)]- 
n_-[E_s(k\!-\!q)]}{q^0+E_r(k)-E_s(k\!-\!q)} - \frac{n_+[E_r(k)]- 
n_+[E_s(k\!-\!q)]}{q^0-E_r(k)+E_s(k\!-\!q)},\\[2ex]
v_{r,s}(k,q) &\equiv& \dsp \frac{1-n_+[E_r(k)]- 
n_-[E_s(k\!-\!q)]}{q^0-E_r(k)-E_s(k\!-\!q)} - \frac{1-n_-[E_r(k)]- 
n_+[E_s(k\!-\!q)]}{q^0+E_r(k)+E_s(k\!-\!q)}, \\[2ex]
&\multicolumn{2}{l}{(r = \pm 1,\quad s=\pm 1)}
\ea
\eeq
For the spatial components, we obtain
\bea
\Pi_{\rm bare}^{ij}&=&  \frac{e^2}{4} \int \frac{d^3k}{(2\pi)^3} \Bigl(\nonumber\\
&&\left\{\de^{ij} \left[1-{\hat k}\cdot(\widehat{k\!-\!q})\right]+ 
\left[{\hat k}^i(\widehat{k\!-\!q})^j + (\widehat{k\!-\!q})^i{\hat k}^j\right]- 
i \eps^{ijl}\left[{\hat k}_l-(\widehat{k\!-\!q})_l\right]\right\}\nonumber\\
&&\qquad \times \Bigl\{\left[1+f_{+,+}^{(2)}(k,q) \right]u_{+,+}(k,q)+ 
\left[1-f_{+,+}^{(2)}(k,q)\right]v_{+,+}(k,q)\Bigr\}
\nonumber\\
&& +\left\{\de^{ij} \left[1-{\hat k}\cdot(\widehat{k\!-\!q})\right]+ 
\left[{\hat k}^i(\widehat{k\!-\!q})^j + (\widehat{k\!-\!q})^i{\hat k}^j\right]+ 
i \eps^{ijl}\left[{\hat k}_l-(\widehat{k\!-\!q})_l\right]\right\}\nonumber\\
&&\qquad \times \Bigl\{\left[1+f_{-,-}^{(2)}(k,q)\right]
u_{-,-}(k,q) + \left[1-f_{-,-}^{(2)}(k,q)\right]v_{-,-}(k,q)\Bigr\}
\nonumber\\
&& +\left\{\de^{ij} \left[1+{\hat k}\cdot(\widehat{k\!-\!q})\right]- 
\left[{\hat k}^i(\widehat{k\!-\!q})^j + (\widehat{k\!-\!q})^i{\hat 
k}^j\right]-i\eps^{ijl}\left[{\hat k}_l+(\widehat{k\!-\!q})_l\right]\right\}\nonumber\\
&&\qquad \times \Bigl\{\left[1+f_{+,-}^{(2)}(k,q)\right]
u_{+,-}(k,q) + \left[1-f_{+,-}^{(2)}(k,q)\right]v_{+,-}(k,q)\Bigr\}
\nonumber\\
&&+\left\{\de^{ij} \left[1+{\hat k}\cdot(\widehat{k\!-\!q})\right]- 
\left[{\hat k}^i(\widehat{k\!-\!q})^j + (\widehat{k\!-\!q})^i{\hat 
k}^j\right]+i\eps^{ijl}\left[{\hat k}_l+(\widehat{k\!-\!q})_l\right]\right\}\nonumber\\
&&\qquad \times \Bigl\{\left[1+f_{-,+}^{(2)}(k,q)\right]u_{-,+}(k,q) + 
\left[1-f_{-,+}^{(2)}(k,q)\right]v_{-,+}(k,q)\Bigr\}\nonumber\\
&& + (\de\mu\to-\de\mu)\Bigr),
\label{Meissner_freqsum}
\eea
where
\beq
f_{r,s}^{(2)}(k,q) \equiv \frac{\left(r|{\vec 
k}|+\mubar\right)\left(s|{\vec k}-{\vec q}|+\mubar\right)-\De^2}{E_r(k)E_s(k\!-\!q)}.
\eeq

Now we can send the external momentum $q\to 0$, to obtain integral
expressions for the Debye and Meissner masses (see section \ref{sec:neutral}).

\subsection{Charged condensate}

In the formalism we are using here, the only difference between the
neutral case and charged case is the sign in front of $\De^2$ in the
numerators. If we change all the signs of $\De^2$ in the numerators in
the previous subsection and leave the denominators unchanged, we will
obtain the correct formulas for the photon self energy in the
presence of a charged condensate.

The zero-zero component of the photon self-energy is
\bea
\Pi_{\rm bare}^{00} 
&=& e^2 \int \frac{d^4k}{(2\pi)^4} \Biggl\{\left[1+{\hat k}\cdot(\widehat{k\!-\!q})\right] 
\frac{(k^0+\dmu)(k^0-q^0+\dmu)+\left(|{\vec k}|+\bar\mu\right)\left(|{\vec k}-{\vec q}|+\bar\mu\right) -\De^2}{D_1(k)D_1(k\!-\!q)}\nonumber\\
&& \qquad +\left[1-{\hat k}\cdot(\widehat{k\!-\!q})\right] 
\frac{(k^0+\dmu)(k^0-q^0+\dmu)-\left(|{\vec k}|+\bar\mu\right)\left(|{\vec k}-{\vec q}|-\bar\mu\right) -\De^2}{D_1(k)D_2(k\!-\!q)}\nonumber\\
&& \qquad + (\dmu\to\dmu)(\bar\mu\to-\bar\mu) + (\dmu\to-\dmu)(\bar\mu\to\bar\mu) + (\dmu\to-\dmu)(\bar\mu\to-\bar\mu)\Biggr\}.
\eea

The spatial components of the photon self-energy are
\bea
 \Pi_{\rm bare}^{ij}&=& e^2 \int \frac{d^4k}{(2\pi)^4} \Biggl\{\nonumber\\
&& \quad \de^{ij} \left[\left(1-{\hat k}\cdot(\widehat{k\!-\!q})\right) 
\frac{(k^0+\dmu)(k^0-q^0+\dmu)+\left(|{\vec k}|+\bar\mu\right)\left(|{\vec k}-{\vec q}|+\bar\mu\right) +\De^2}{D_1(k)D_1(k\!-\!q)}\right.\nonumber\\
&& \qquad \left.+\left(1+{\hat k}\cdot(\widehat{k\!-\!q})\right)
\frac{(k^0+\dmu)(k^0-q^0+\dmu)-\left(|{\vec k}|+\bar\mu\right)\left(|{\vec k}-{\vec q}|-\bar\mu\right) +\De^2}{D_1(k)D_2(k\!-\!q)}\right]\nonumber\\
&& \quad +\left[{\hat k}^i(\widehat{k\!-\!q})^j + (\widehat{k\!-\!q})^i{\hat k}^j\right]\left[\frac{(k^0+\dmu)(k^0-q^0+\dmu)+\left(|{\vec k}|+\bar\mu\right)\left(|{\vec k}-{\vec q}|+\bar\mu\right) +\De^2}{D_1(k)D_1(k\!-\!q)}\right.\nonumber\\
&& \qquad \left.-\frac{(k^0+\dmu)(k^0-q^0+\dmu)-\left(|{\vec k}|+\bar\mu\right)\left(|{\vec k}-{\vec q}|-\bar\mu\right) +\De^2}{D_1(k)D_2(k\!-\!q)}\right]\nonumber\\
&& \quad +(\bar\mu\to-\bar\mu)\nonumber\\
&& \quad + i \eps^{ijl}\Biggl[\left({\hat k}_l-(\widehat {k\!-\!q})_l\right)\left(-\frac{(k^0+\dmu)(k^0-q^0+\dmu)+\left(|{\vec k}|+\bar\mu\right)\left(|{\vec k}-{\vec q}|+\bar\mu\right) +\De^2}{D_1(k)D_1(k\!-\!q)}\right.\nonumber\\
&& \qquad \quad \left. + \frac{(k^0+\dmu)(k^0-q^0+\dmu)+\left(|{\vec k}|-\bar\mu\right)\left(|{\vec k}-{\vec q}|-\bar\mu\right) +\De^2}{D_2(k)D_2(k\!-\!q)}\right)\nonumber\\
&&\qquad +\left({\hat k}_l+(\widehat {k\!-\!q})_l\right)\left(-\frac{(k^0+\dmu)(k^0-q^0+\dmu)-\left(|{\vec k}|+\bar\mu\right)\left(|{\vec k}-{\vec q}|+\bar\mu\right) +\De^2}{D_1(k)D_2(k\!-\!q)}\right.\nonumber\\
&& \qquad \quad\left. + \frac{(k^0+\dmu)(k^0-q^0+\dmu)-\left(|{\vec k}|-\bar\mu\right)\left(|{\vec k}-{\vec q}|-\bar\mu\right) +\De^2}{D_2(k)D_1(k\!-\!q)}\right)\Biggr]\nonumber\\
&& \quad +(\dmu\to-\dmu)\Biggr\}.
\eea

To perform the frequency sum we can take the neutral-condensate
results and interchange $f_{r,s}^{(1)}(k,p)$ and $f_{r,s}^{(2)}(k,p)$. 
Thus we obtain
\beq
\ba{rl}
\dsp \Pi_{\rm bare}^{00} = \frac{e^2}{4} \int \frac{d^3k}{(2\pi)^3} \Bigl\{
& \left[1+{\hat k}\cdot(\widehat{k\!-\!q})\right]
  \left[1+f_{+,+}^{(2)}(k,p)\right]u_{+,+}(k,p)\\[1ex]
 + & \left[1+{\hat k}\cdot(\widehat{k\!-\!q})\right]
   \left[1-f_{+,+}^{(2)}(k,p)\right]v_{+,+}(k,p)\\[1ex]
 + & \left[1-{\hat k}\cdot(\widehat{k\!-\!q})\right] 
   \left[1+f_{+,-}^{(2)}(k,p)\right]u_{+,-}(k,p)\\[1ex]
 + & \left[1-{\hat k}\cdot(\widehat{k\!-\!q})\right]
   \left[1-f_{+,-}^{(2)}(k,p)\right]v_{+,-}(k,p)\\
 + &  (\de\mu\to\de\mu)(\bar\mu\to-\bar\mu) + (\de\mu\to-\de\mu)(\bar\mu\to\bar\mu) \\
 + &  (\de\mu\to-\de\mu)(\bar\mu\to-\bar\mu)\Bigr\},
\ea
\label{charged_Debye_freqsum}
\eeq
and 
\bea
\Pi_{\rm bare}^{ij}&=&  \frac{e^2}{4} \int \frac{d^3k}{(2\pi)^3} \Bigl(\nonumber\\
&&\left\{\de^{ij} \left[1-{\hat k}\cdot(\widehat{k\!-\!q})\right]+ 
\left[{\hat k}^i(\widehat{k\!-\!q})^j + (\widehat{k\!-\!q})^i{\hat k}^j\right]- 
i \eps^{ijl}\left[{\hat k}_l-(\widehat{k\!-\!q})_l\right]\right\}\nonumber\\
&&\qquad \times \Bigl\{\left[1+f_{+,+}^{(1)}(k,q) \right]u_{+,+}(k,q)+ 
\left[1-f_{+,+}^{(1)}(k,q)\right]v_{+,+}(k,q)\Bigr\}
\nonumber\\
&& +\left\{\de^{ij} \left[1-{\hat k}\cdot(\widehat{k\!-\!q})\right]+ 
\left[{\hat k}^i(\widehat{k\!-\!q})^j + (\widehat{k\!-\!q})^i{\hat k}^j\right]+ 
i \eps^{ijl}\left[{\hat k}_l-(\widehat{k\!-\!q})_l\right]\right\}\nonumber\\
&&\qquad \times \Bigl\{\left[1+f_{-,-}^{(1)}(k,q)\right]
u_{-,-}(k,q) + \left[1-f_{-,-}^{(1)}(k,q)\right]v_{-,-}(k,q)\Bigr\}
\nonumber\\
&& +\left\{\de^{ij} \left[1+{\hat k}\cdot(\widehat{k\!-\!q})\right]- 
\left[{\hat k}^i(\widehat{k\!-\!q})^j + (\widehat{k\!-\!q})^i{\hat 
k}^j\right]-i\eps^{ijl}\left[{\hat k}_l+(\widehat{k\!-\!q})_l\right]\right\}\nonumber\\
&&\qquad \times \Bigl\{\left[1+f_{+,-}^{(1)}(k,q)\right]
u_{+,-}(k,q) + \left[1-f_{+,-}^{(1)}(k,q)\right]v_{+,-}(k,q)\Bigr\}
\nonumber\\
&&+\left\{\de^{ij} \left[1+{\hat k}\cdot(\widehat{k\!-\!q})\right]- 
\left[{\hat k}^i(\widehat{k\!-\!q})^j + (\widehat{k\!-\!q})^i{\hat 
k}^j\right]+i\eps^{ijl}\left[{\hat k}_l+(\widehat{k\!-\!q})_l\right]\right\}\nonumber\\
&&\qquad \times \Bigl\{\left[1+f_{-,+}^{(1)}(k,q)\right]u_{-,+}(k,q) + 
\left[1-f_{-,+}^{(1)}(k,q)\right]v_{-,+}(k,q)\Bigr\}\nonumber\\
&& + (\de\mu\to-\de\mu)\Bigr).
\label{charged_Meissner_freqsum}
\eea
Now we can send the external momentum $q\to 0$, to obtain integral
expressions for the Debye and Meissner masses (see section \ref{sec:charged}).

\end{document}